# Financial Studio: Android Based Application for Computing Tax, Pension, Zakat and Loan

***Muhammad Zubair Asghar, Ulfat Batool, Farheen Bibi, Sadia Ismail, Syeda Rabail Zahra**

Institute of Computing and Information Technology, Gomal University, D.I.Khan, Pakistan

*Contact author

***Abstract***

*This work deals with the development of android-based financial studio, an integrated application for calculating tax, pension, zakat, and loan. Financial studio can facilitate employers of any department and other individuals. The application is developed using MIT app inventor-based android platform. The financial studio has four computational modules, namely: (i) tax, (ii) pension, (iii) zakat, and (iv) loan. The system provides an integrated environment for performing aforementioned distinct calculations by integrating different financial modules into a single application in a user-friendly way. The statistical analysis shows that the application is effective to deal with different financial calculations.*

## 1. Introduction

Android-based applications are getting popular due to revolution of smart phones and internet. People from every walk of life are now relying on applications stored in their handheld devices.There exists number of applications of Google Play Store, which assist the users for accomplishing different tasks. As far as financial computations are concerned, people are in need of such commonly used applications which can calculate tax, loan, zakat and pension. Therefore, it is an important task to develop an android-based application that can facilitate employers and other individuals in a better way to calculate their income taxes, pension, zakat, and loan amount in a single application on android phone.

A financial studio is an android-based suit of applications to perform different financial computations, such as income tax, loan, zakat, and pension, which are commonly needed. It facilitates employers of any department and other individuals. It also assists users to assess, how much, one should invest every month to get a desired amount at the end of an investment. Financial studio application can quickly calculate taxes, retirement plans, pension, loans, financial investments, and zakat. . The existing android applications lack many important features needed on a daily basis. This is what we address in this work.
In literature, there exists number of financial applications. In the following paragraphs, we present some selected studies conducted for the development of such systems.

A web based tax calculation system is proposed by [1], which applies income tax rates in Pakistan on taxable income of salaried persons and salaried class. A salaried person class is applicable, where the salary income exceeds than the 50% of the total taxable income. A web based Pension calculator is developed for Government of Khyber Pakhtunkhwa Finance Department [3]. It is very simple calculator and requires some inputs to calculate pension. We can develop an improved version of such calculator by deploying it on android device to calculate the pension of pensioner of any department.

The web based zakat calculator [5] is used for calculation of zakat based on 2.5% of the total net worth that is available at the end of one lunar year. The loan calculator [7] helps in calculating the monthly payments on a loan. Users have to simply enter the loan amount, term and interest rate in the fields to calculate. This system can be used for mortgage, auto, or any other fixed loan types. We can create an android driven application of loan calculator that calculate both monthly and yearly loan amount.

Pakistan Income Tax Calculator [10] is very useful for salaried people to compute their annual income tax and for tax filing. It is very simple and easy to use. We can customize the tax slab. The pension calculator [14] application allows public sector workers to calculate the pension entitlements, they will receive from their superannuation scheme at retirement, based on their years' service. This application also allows us to calculate what potential extra retirement fund we could accumulate by contributing to an AVC today.

Zakat is one of the five pillars of Islam. The avis Team developed an android application [16] to assist in calculating Zakat by making the zakat payables easy, and accurate. Moreover, to calculate loan, an easy to use system is developed by [17] on the basis of annuity, differentiated, and fixed payments calculation. There are many systems on the Financial-calculators content analysis in the context of software engineering, opinion mining and sentiment analysis [18, 19, 20, 21, 22, 25, 26, 27], however, most of such studies are web-based and address the user generated contents.

The above financial calculator systems are either web-based recommendation systems or intelligent expert systems. Moreover, the above mentioned systems are not available in single android application and lack most of the important features. Therefore, there is a need to develop an android-based easy to use application that can assist the users to monitor their financial management activities. This what we will address in this work.

## 2. Methodology

The proposed system is comprised of four modules, namely: (i) tax calculation, (ii) pension calculation, (iii) zakat calculation, and (iv) tax calculation. The flowchart representation of the proposed system is shown in Fig. 1.

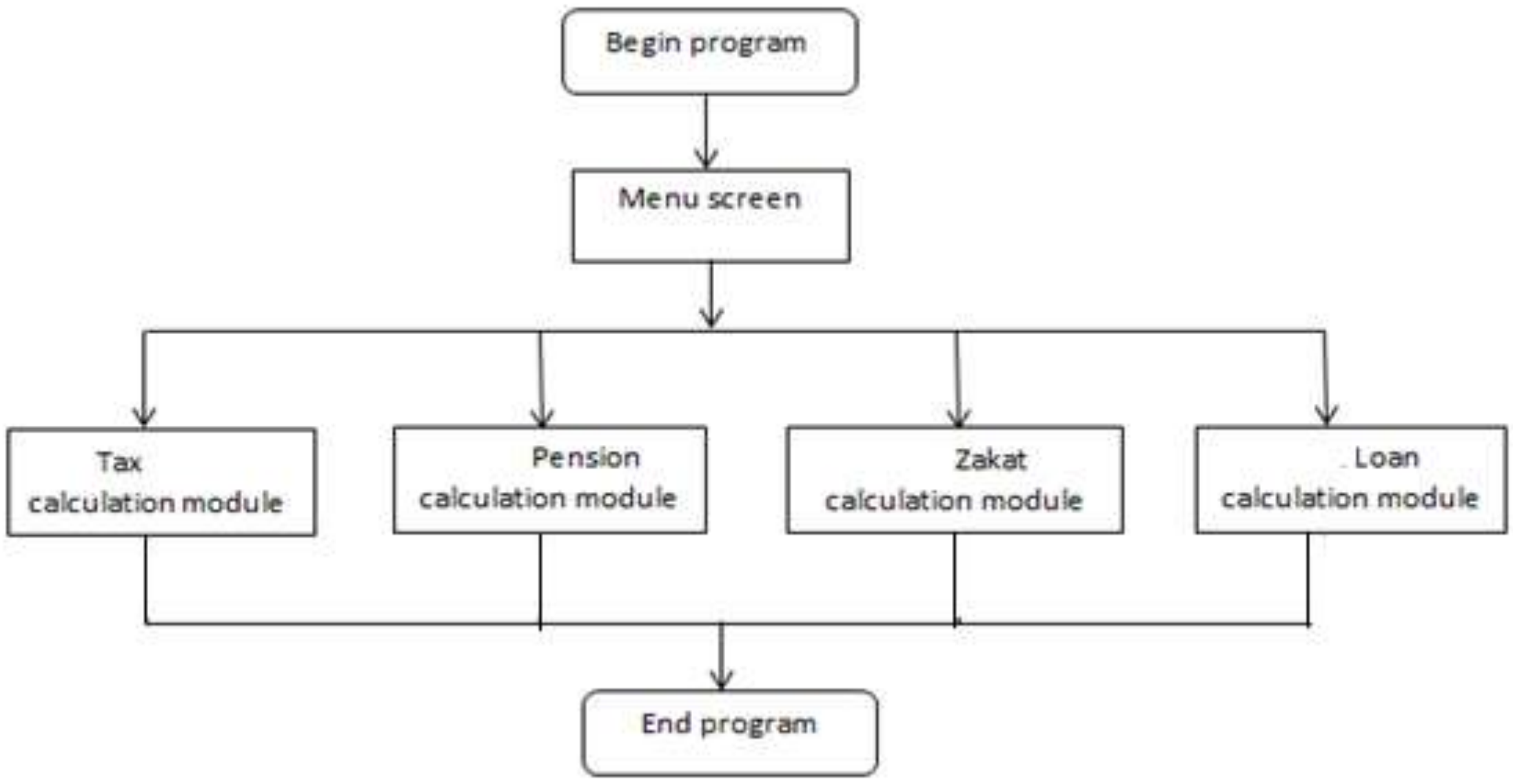

**Fig. 1** Flowchart of the proposed system

**2.1** ***Tax calculation module:*** This module deals with the calculation of income tax for salaried persons. We used different labels and text boxes, such as name, CNIC, NTN, designation, posting city, employer NTN, Tax year, date, and monthly income for entering required data of employees to be used for the calculations of tax. There are three buttons: already paid tax button, teacher exemption button, and calculate tax button. If users have already paid any tax then clicking on "*already paid tax button*", provides a drop down list of textboxes, and user fills out these boxes. If the user is a teacher, then, upon clicking on "*teacher exemption button*", a checkbox appears, which, if checked, provides 40% exemption of the tax. The "*calculate tax button*" enables the user click to display result. The flowchart of tax calculation module is shown in Fig. 2

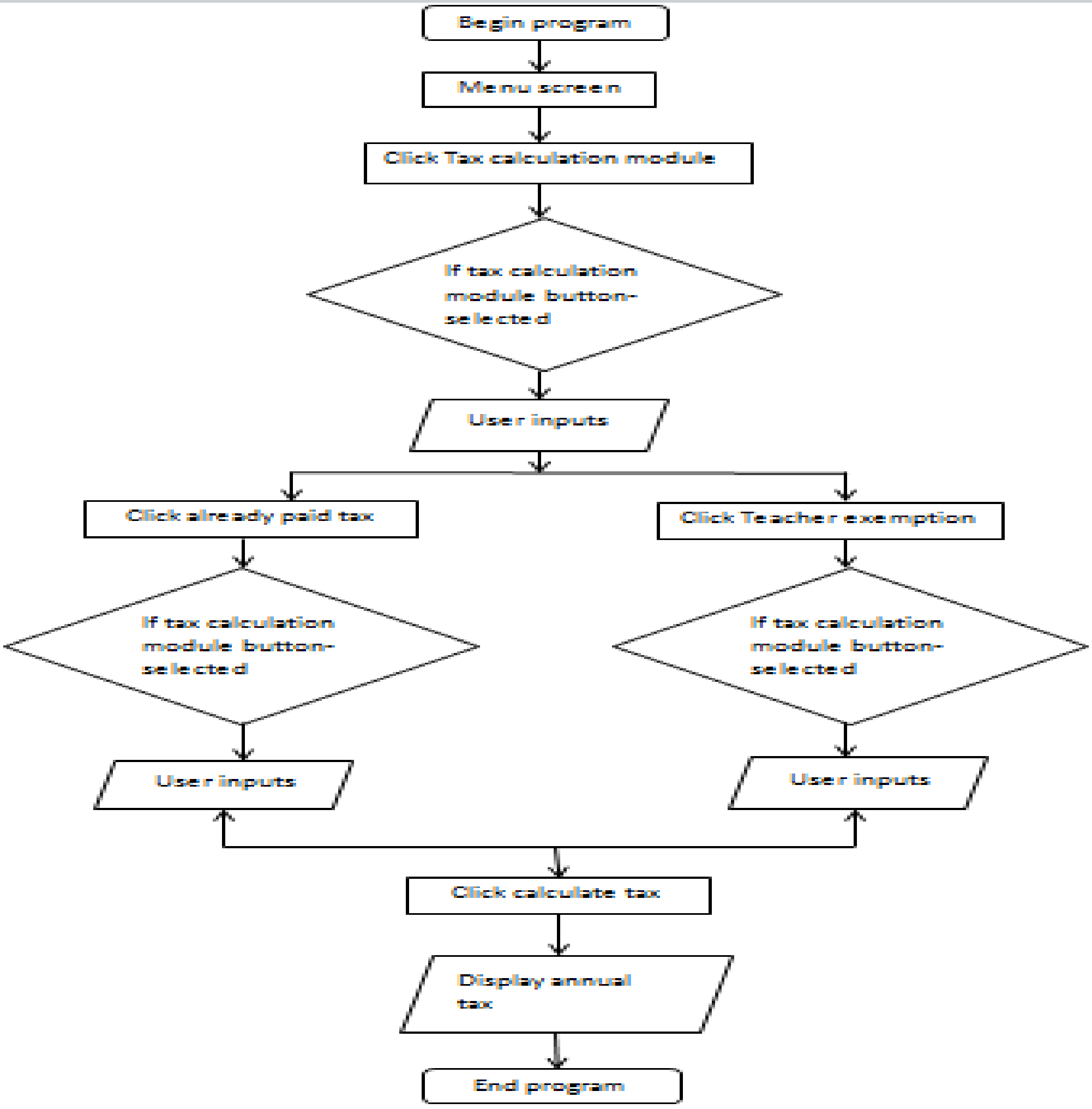

**Fig. 2.** Tax calculation Flow Chart

2.2 ***Pension calculation module:*** The pension calculation aims at computing the amount of pension, gratuity and increases to be received by a retired government employee at the age of superannuation (60 years alive). The user makes inputs like name of pensioner date of birth, date of appointment, date of retirement, BPS, last basic pay, qualifying service using text boxes and labels. When all of the required inputs are completed then "*calculate pension button*" is used to display the results. The flowchart of pension calculation module is shown in Fig. 3 and the pseudo code algorithm is shown in Algorithm 1.

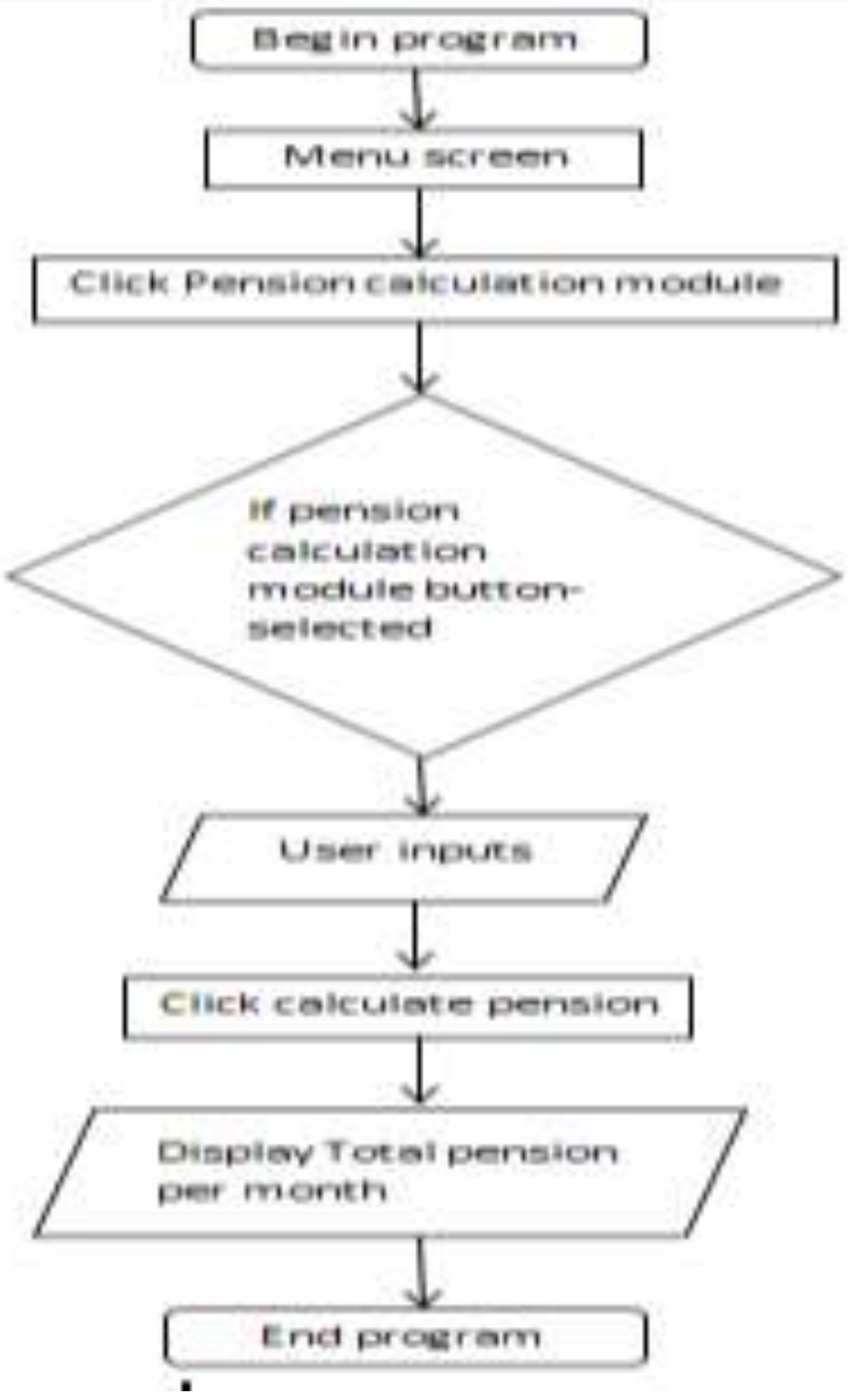

Fig. 3 Pension calculation flowchart

**Algorithm 1.** Pension Calculation module

**Input:** Name of pensioner, Date of birth, Date of appointment, Date of retirement, BPS, Last basic pay, qualifying service
**Output:** Net pension, Total gratuity, increases 2010 to 2015, 25% medical allowance of 2010, Total pension per month
**Begin:**
// initializations
TService ←Grasspension ←Comutedportion←0

1. If ( Calculate pension button is clicked ) then
2. If( Qualifying service textbox.Text >= 30) then
3. {
4. Set global Tservice ←30
5. Else Set global Tservice ← Qualifying service textbox. Text
6. Set global Grass pension←last basic pay textbox. Text *7*get global TService/300, Set global commuted portion ← get global Grass pension *35/300, Set Net pension label. Text←get global Grass pension – get global commuted portion, Set Total Gratuity label. Text ←get global commuted portion * 148.4628, Set AR 2010 label. Text ← Net pension label. Text * 15/100, Set AR 2011 label. Text ← Net pension label. Text * 15/100, Set AR 2012 label. Text ← Net pension label. Text * 20/100, Set AR 2013 label. Text ←Net pension label. Text * 15/100,Set AR 2014 label. Text ← Net pension label. Text * 10/100, Set AR 2015 label. Text ←Net pension label. Text * 10/100, Set MA 2010 label. Text ←Net pension label. Text * 25/100, Set Total pension per month label. Text =AR 2010 label. Text + AR 2011 label. Text + AR 2012 label. Text + AR 2013 label.Text+ AR 2014 label. Text + AR 2015 label. Text + MA2010 label. Text + Net pension label. Text
7. }
8. **End**

---

2.3 ***Zakat calculation module:*** The zakat calculation module based on Islamic law, enables the user to calculate the zakat payable by the user based on 2.5% of the total worth of the assets. This module facilitates the user in computing the Zakat to be paid on the basis of: (i) gold, (ii) silver, (iii) cash, (iv) business, and (v) property. The user first has to choose one from five types by clicking on the corresponding button. Finally, "*calculate Zakat button*" is used to display the results. Pseudo code steps for zakat calculation module are presented in Algorithm 2.

**Algorithm 2.** Zakat Calculation Module

**Input:** Gold, Silver, Cash, Business, Property.
**Output:** Total assets, Zakat due.
**Begin:**
// initializations
Gold← Silver← Cash← Business ←Property← Total price← 0

1. If (Calculate Zakat button is clicked) then
2. If( WOG Textbox. Text>=7.5) then
3. Set global Total price←WOG textbox. Text * POG textbox. Text,Set global Gold←get global total price
4. If (WOS textbox. Text>=52.5) then
5. Set global Total price←WOS textbox. Text * POS textbox. Text, Set global Silver←get global total price
6. Set global Total price←CHB textbox. Text + BAS textbox. Text+ SSS textbox. Text + MYHL textbox+ OCM textbox. Text
7. If (global Total price>= textbox. Text ) then
8. Set global Cash←get global total price, Set global Total price←BI textbox. Text + PFI textbox. Text + BS textbox. Text + OFB text, Set global Business← get global total price, Set global Total price ←Net property textbox. Text + Other property textbox. Text
9. If (get global Total price >=PAZ textbox. Text) then
10. Set global property ←get global total price, Set Total assets label.text ←get global Gold+ get global Silver +get global Cash+ get global Busines +get global Property,Set Zakat due label.text ←Total assets label.text *2.5/100
11. }
12. **End**

**2.4** ***Loan calculation module:*** This module enables the users to determine the monthly and yearly payments on the loan they acquired. The user simply enters the loan amount, term and interest rate in the text fields and click "*calculate button*" and get the desired output. . The pseudo code algorithm is shown in algorithm 3.

**Algorithm 3.** Loan Calculation Module

**Input:** Loan amount, Rate of interest (Annually) in %, Number of months/years.
**Output:** Monthly payment Amount, Yearly payment Amount
**Begin:**
// initializations
LAmount← ROI ←NOMY←0

1. If (Calculate loan button is clicked) then
2. If( LAmount Textbox.Text= number ) then
3. Set LAmount←L Amount Textbox
4. If (Rate of interest Textbox. Text= number) then
5. Set Rate of interest ←Rate of interest Textbox.text
6. If (Name of months /year Textbox=Number) then
7. Set Name of months /year←Name of months/year Textbox.text, Set Monthly amount label .text← ((get Loan amount* get Rate of interest/100) /12*get number of months/year), Set Yearly amount label .text← ((get Loan amount* get Rate of interest/100) *get number of months/year)
8. }
9. **End**

### 3. Implementation

We executed our financial application namely, tax, pension, zakat, and loan using android based platform. It assists different users including employees and others to perform the required financial calculations on their hand held device easily. Visual block programming language is used for the development of application. The software and hardware resources used to develop the proposed system are as follows: (i) Windows 7 HP laptop, (ii) MIT App inventor 2, (iii) Bluestack emulator, (iv) HUAWEI y6 Android Cell Phone, and (v) Samsung Tablet.

In the following sub-sections, we present the output screen and partial list of coding of different modules of the proposed system.

### 3.1 Main Module

An output screen and a partial list of coding for main module are presented in Fig. 4 and Fig.5 respectively.

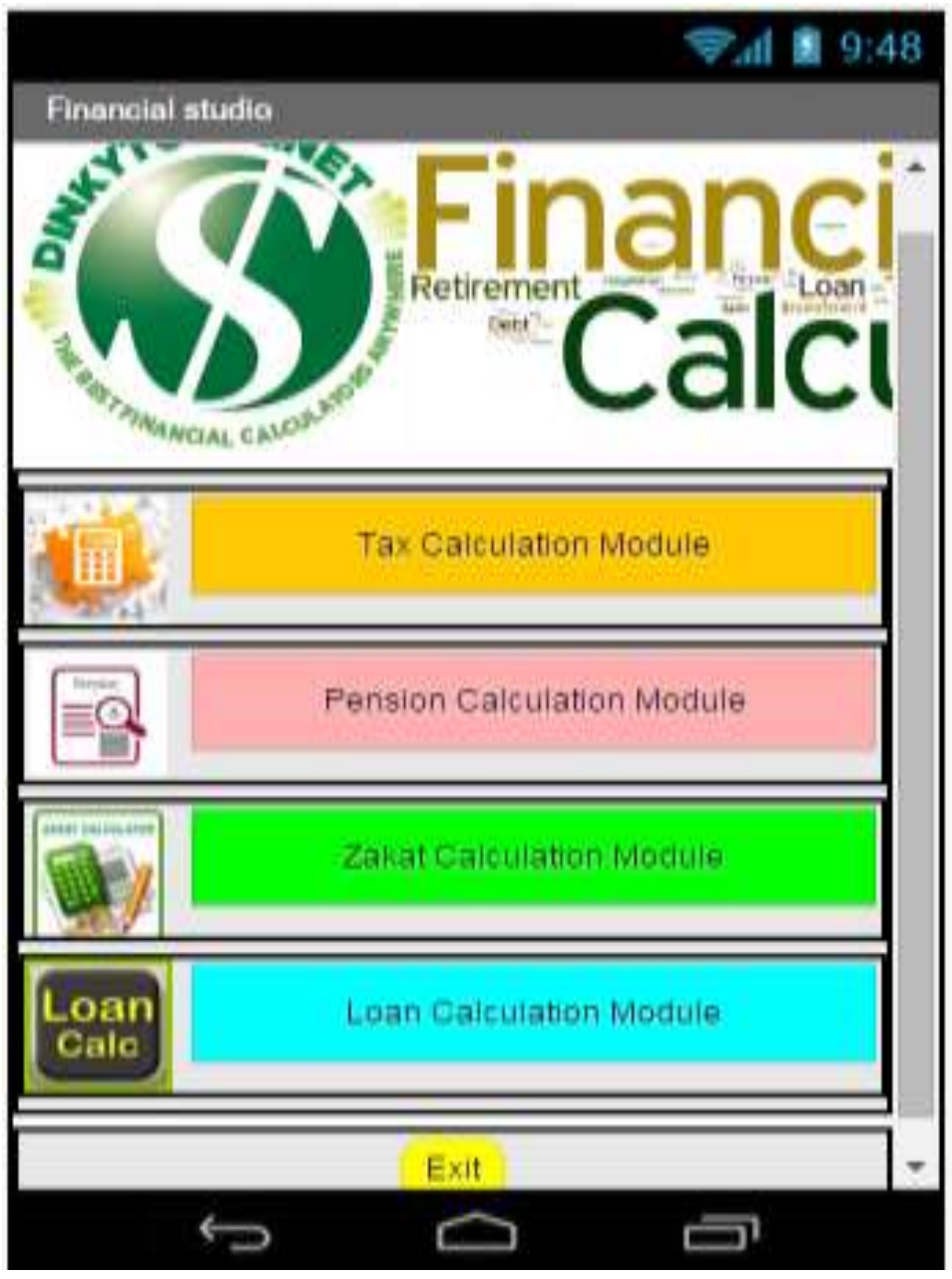

**Fig.4** Output screen of main menu

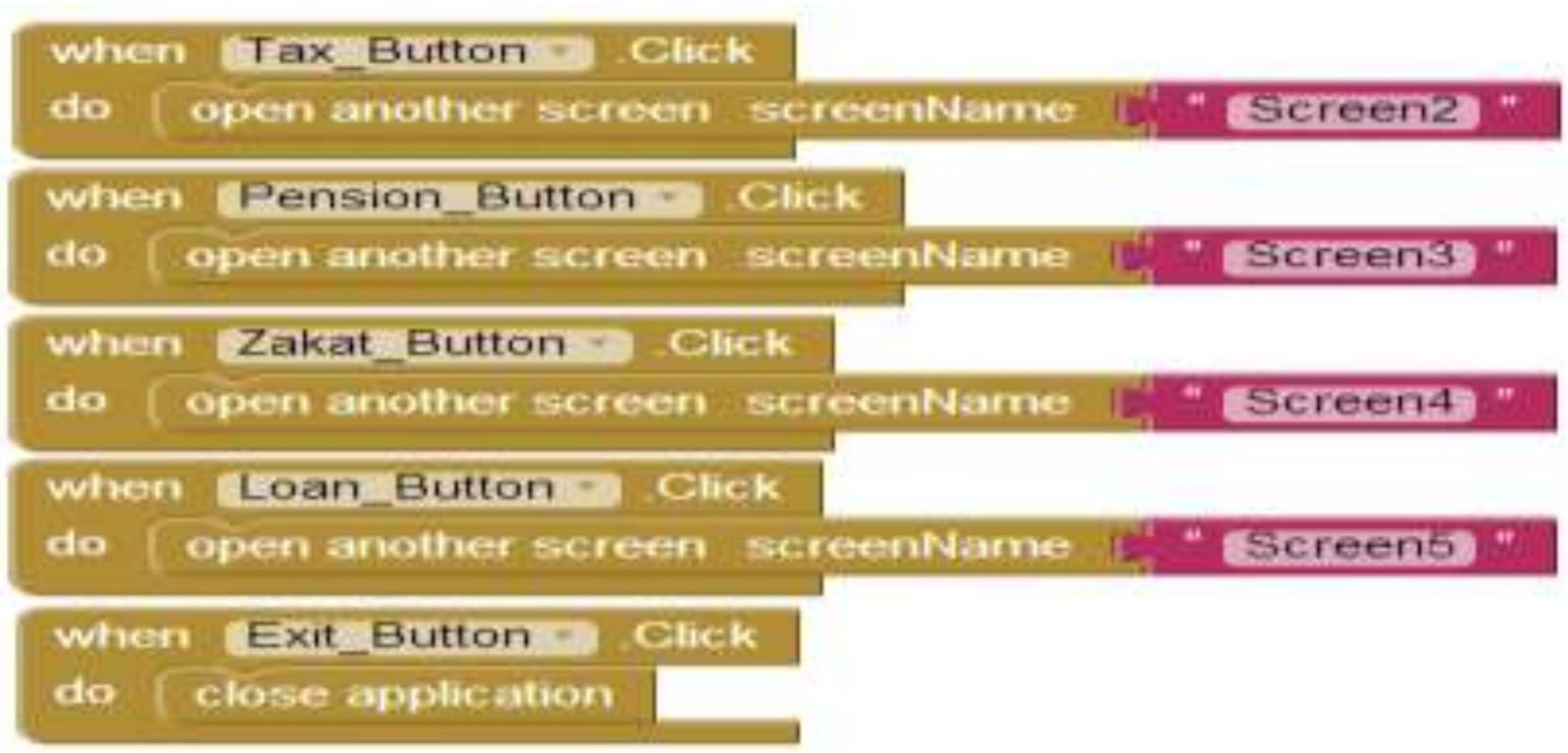

**Fig. 5.** Code block for main module of Financial studio

### 3.2 Tax Calculation Module

An output screen and a partial list of coding for tax calculation module are presented in Fig. 6 and Fig.7 respectively.

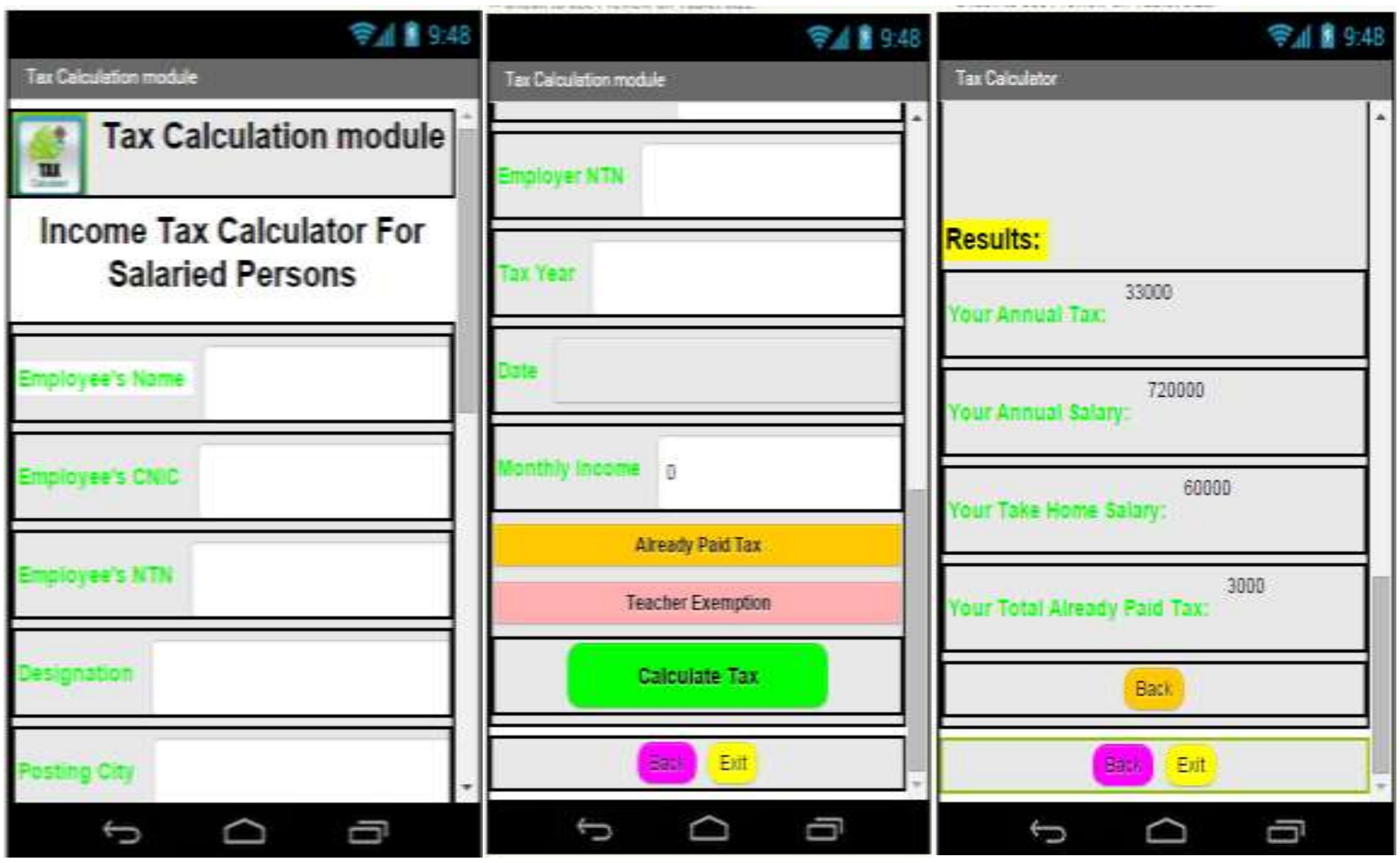

(a) (b) (c)

**Fig. 6.** Tax Calculation module screen (a) input1 (b) input2 (c) output

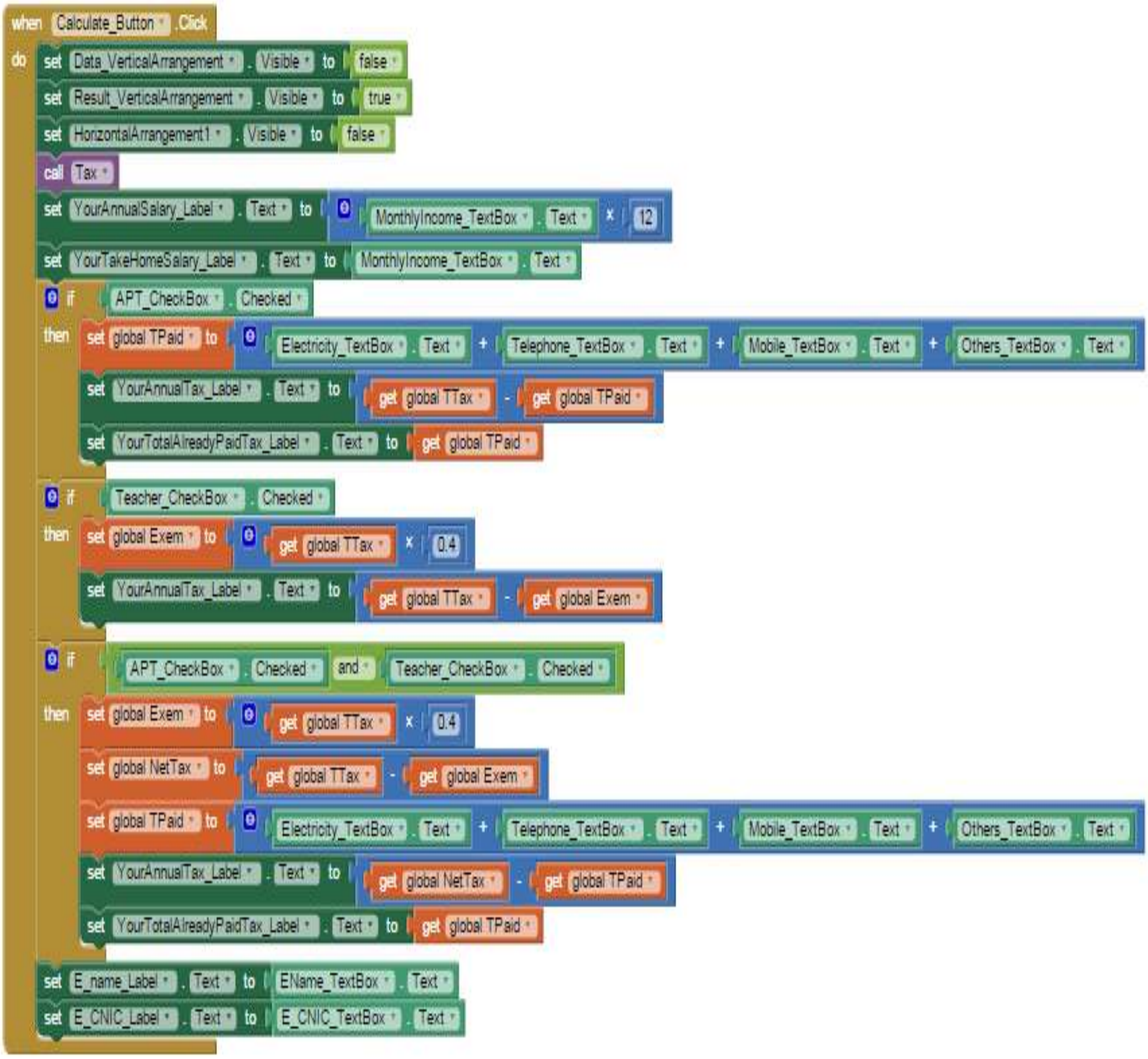

**Fig. 7.** Code block for input and output of Tax calculation module

### 3.3 Pension Calculation Module

An output screen and a partial list of coding for pension calculation module are presented in Fig. 8 and Fig 9 respectively.

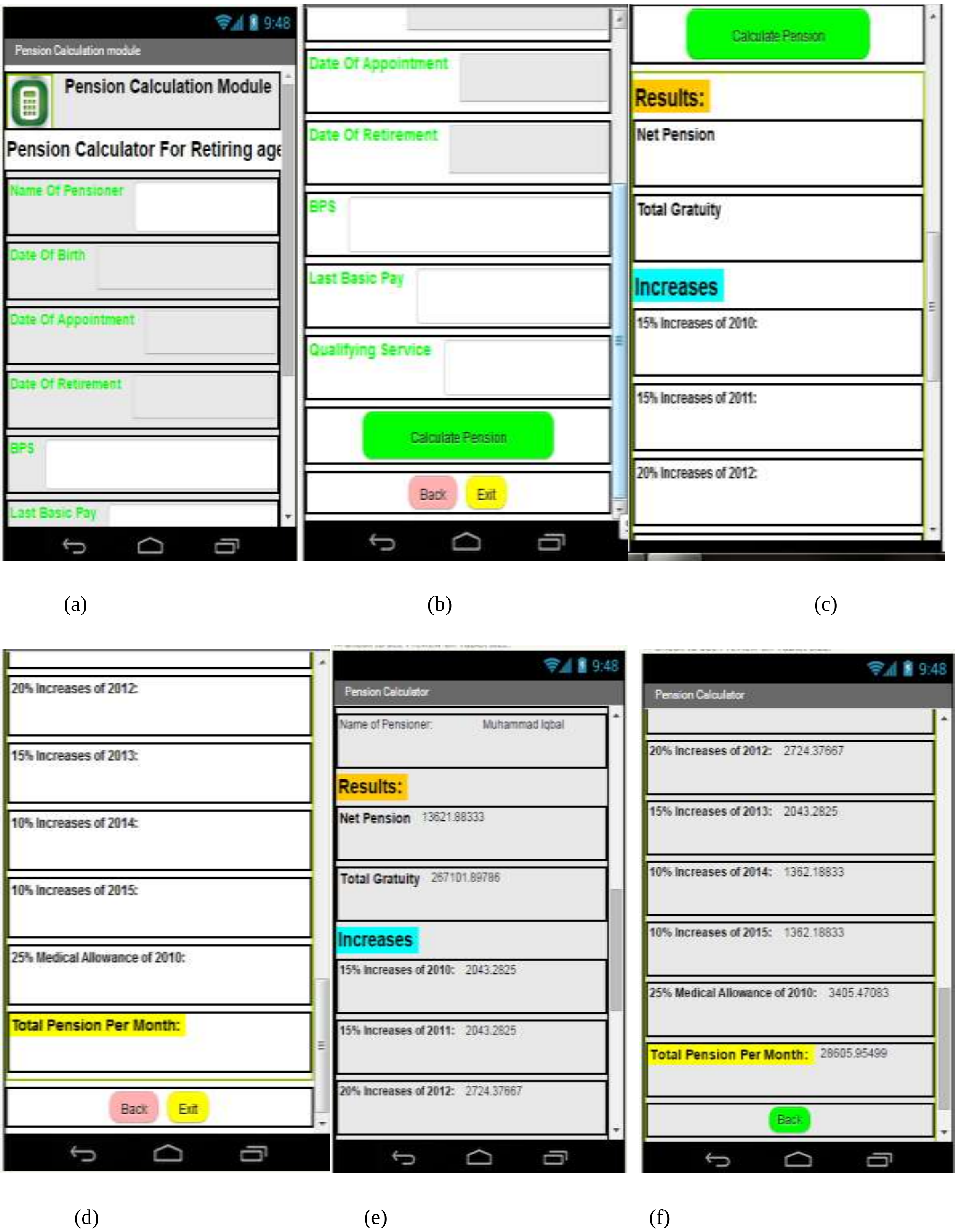

**Fig. 8.** Pension Calculation module screen (a) input1 (b) input2 (c) input3 (d) input4 (e) Output (f) output

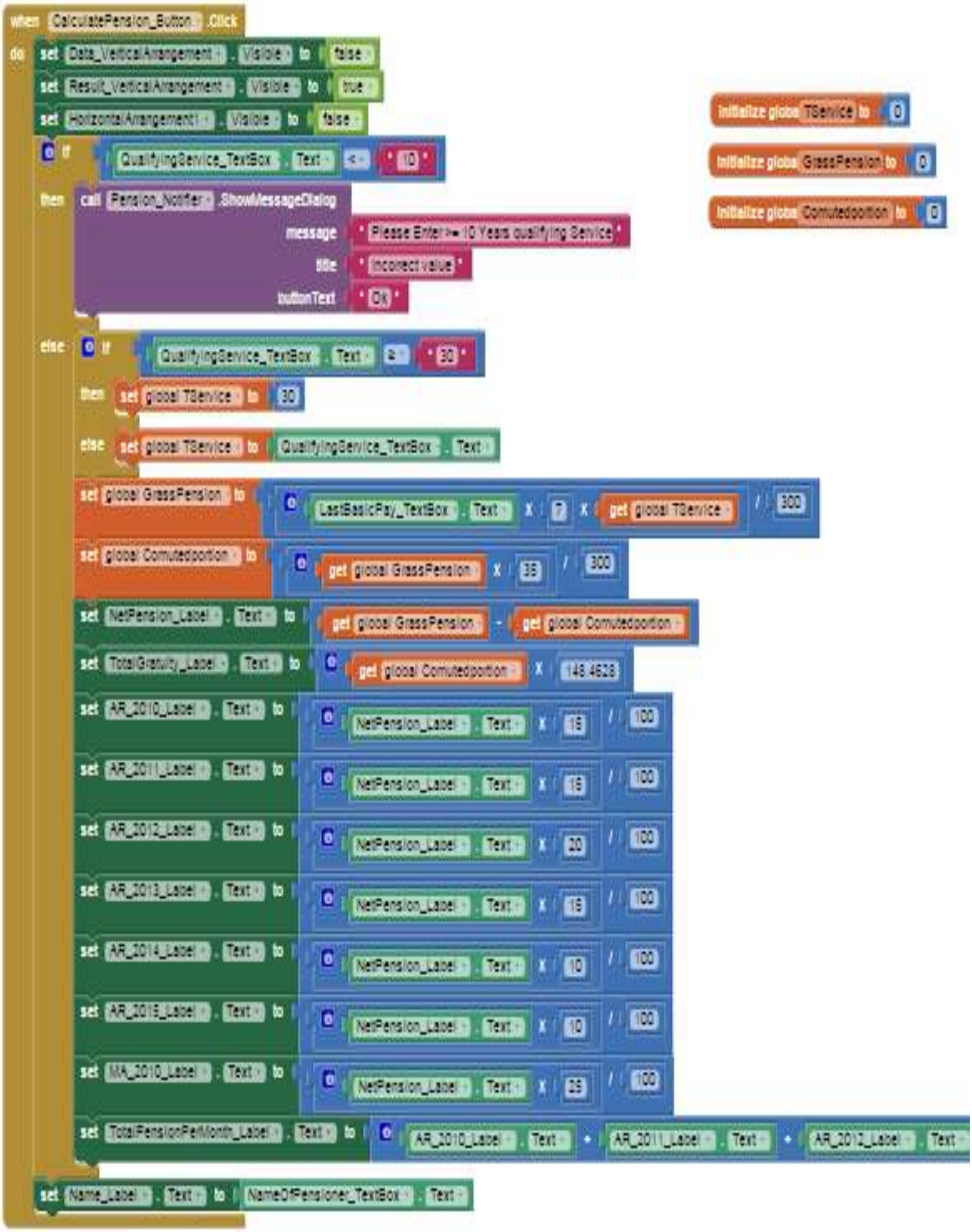

**Fig. 9.** Code block for input and output of Pension calculation module

### 3.4 Zakat Calculation Module

An output screen and a partial list of coding for zakat calculation module are presented in Fig. 10 and Fig.11 respectively.

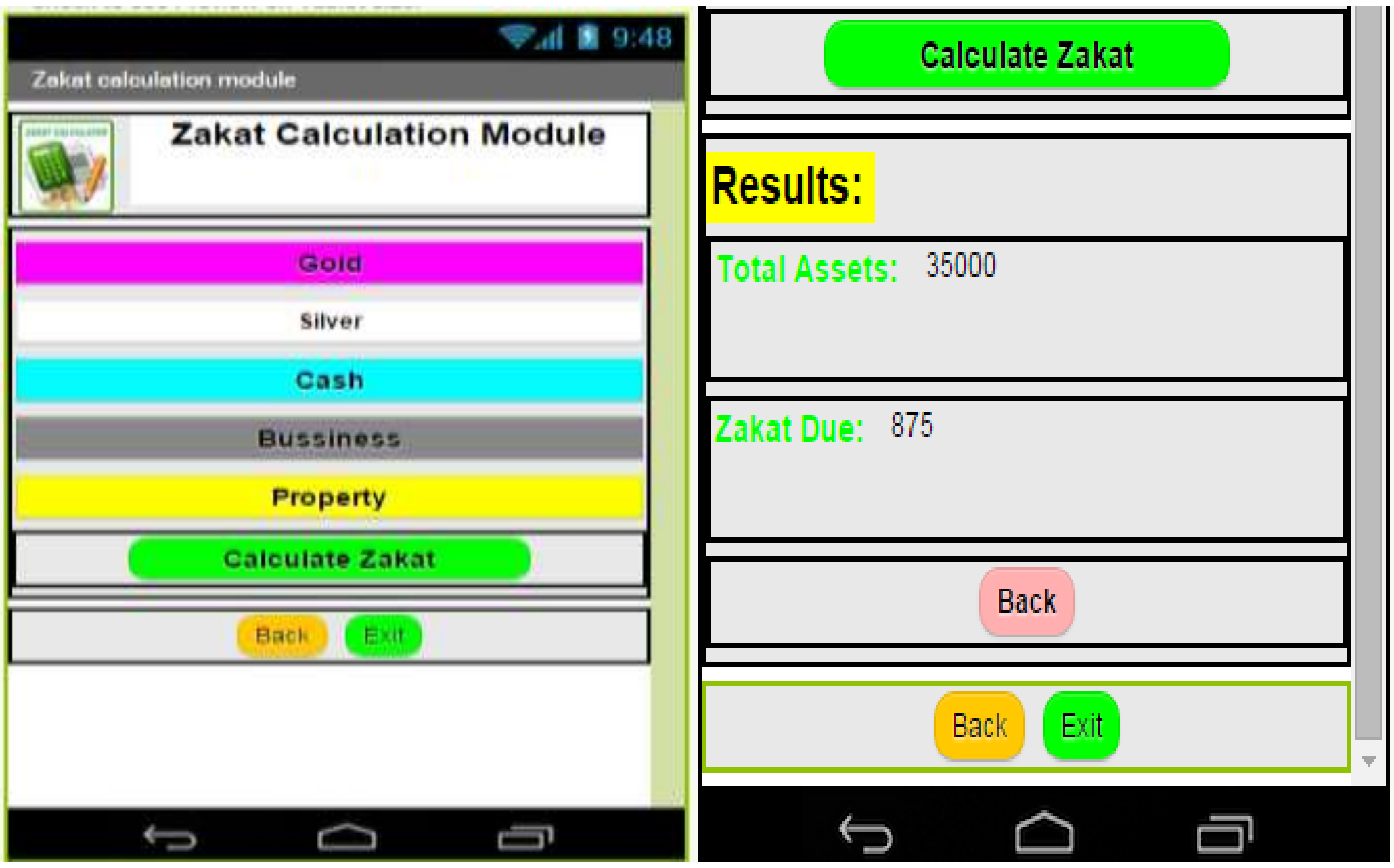

(a) (b)

**Fig. 10.** Zakat Calculation module screen (a) input1 (b) output

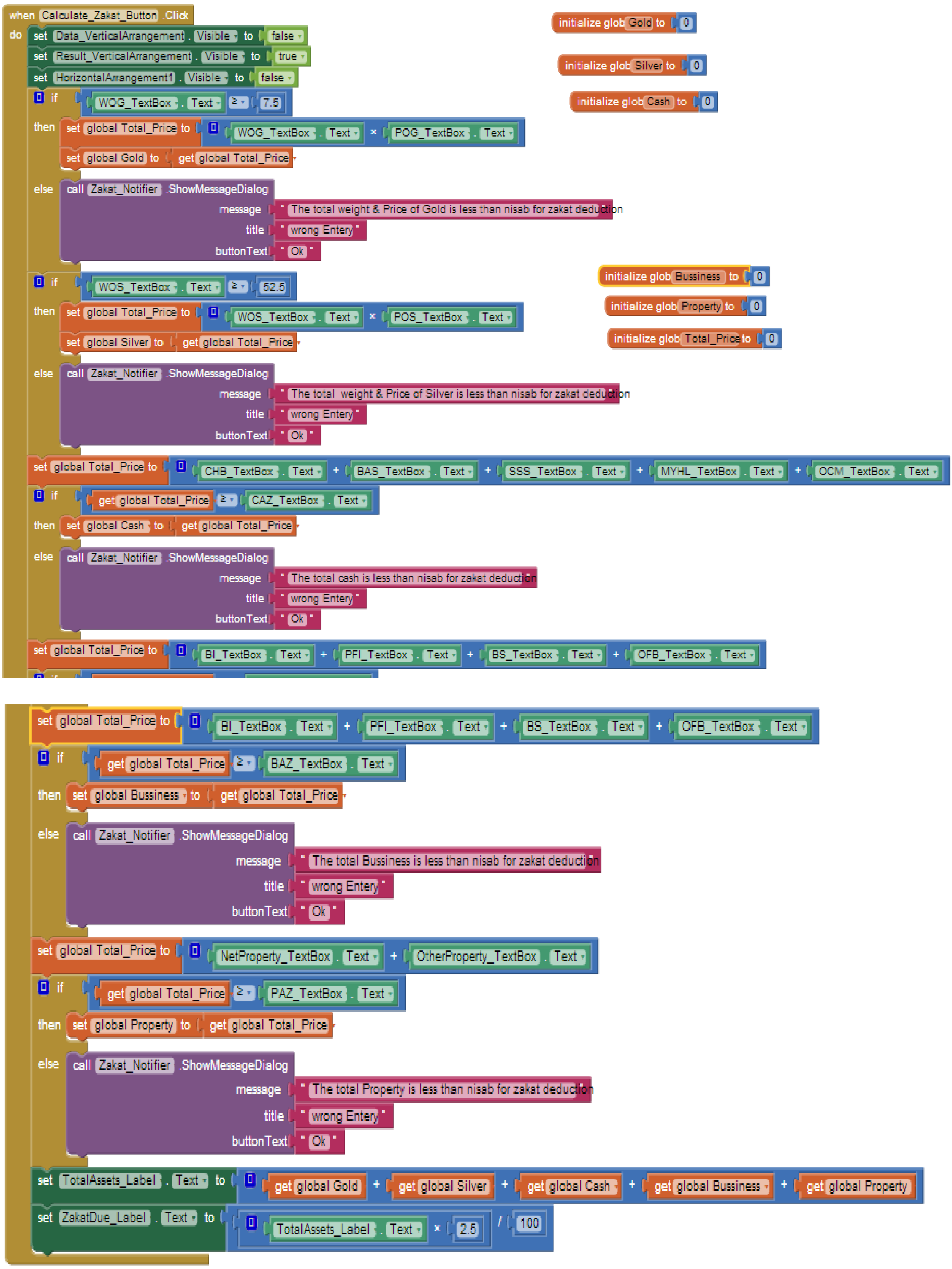

**Fig. 11.** Code block for input and output of Zakat calculation module

### 3.5 Loan Calculation Module

An output screen and a partial list of coding for loan calculation module are presented in Fig. 12 and Fig.13 respectively

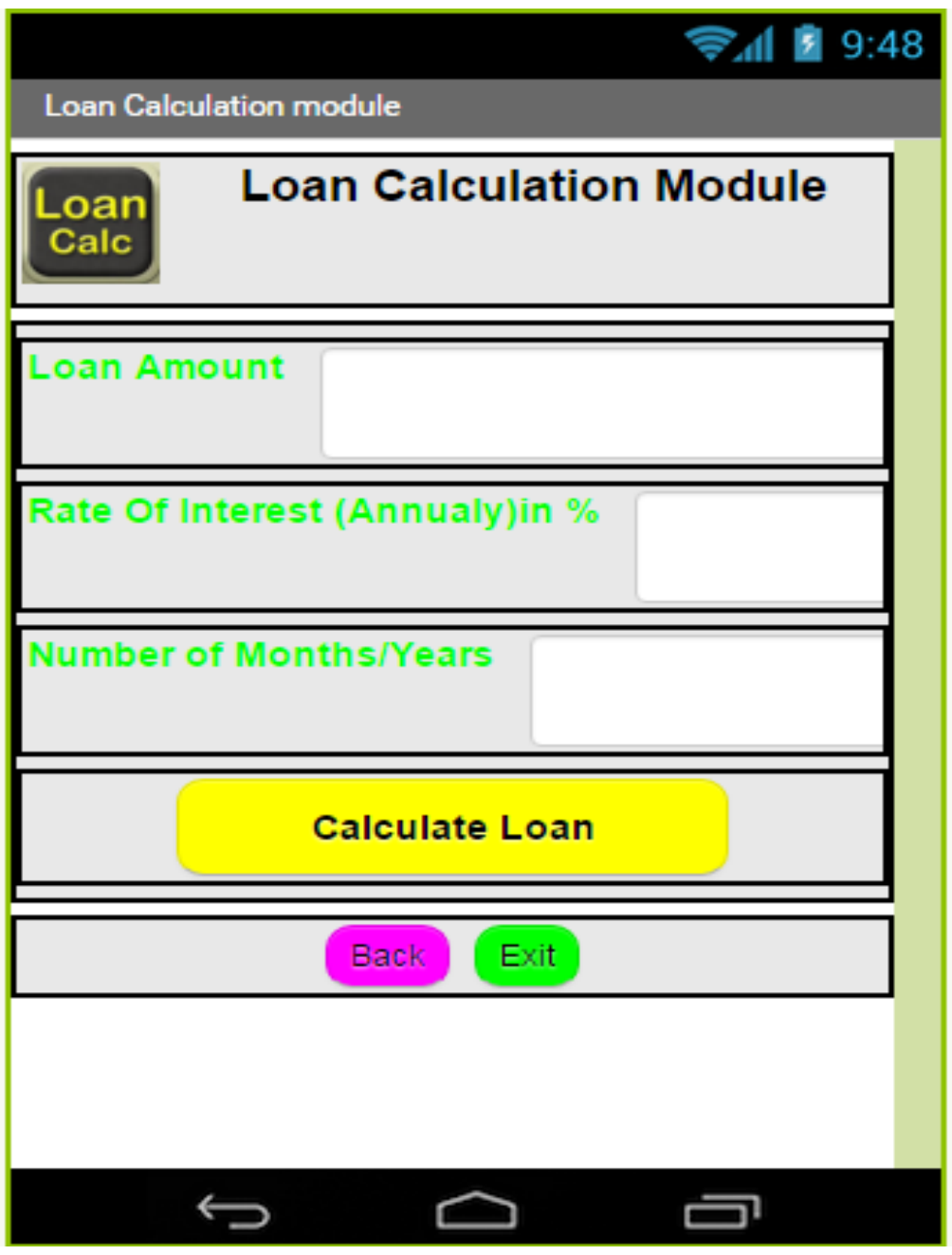

(a)

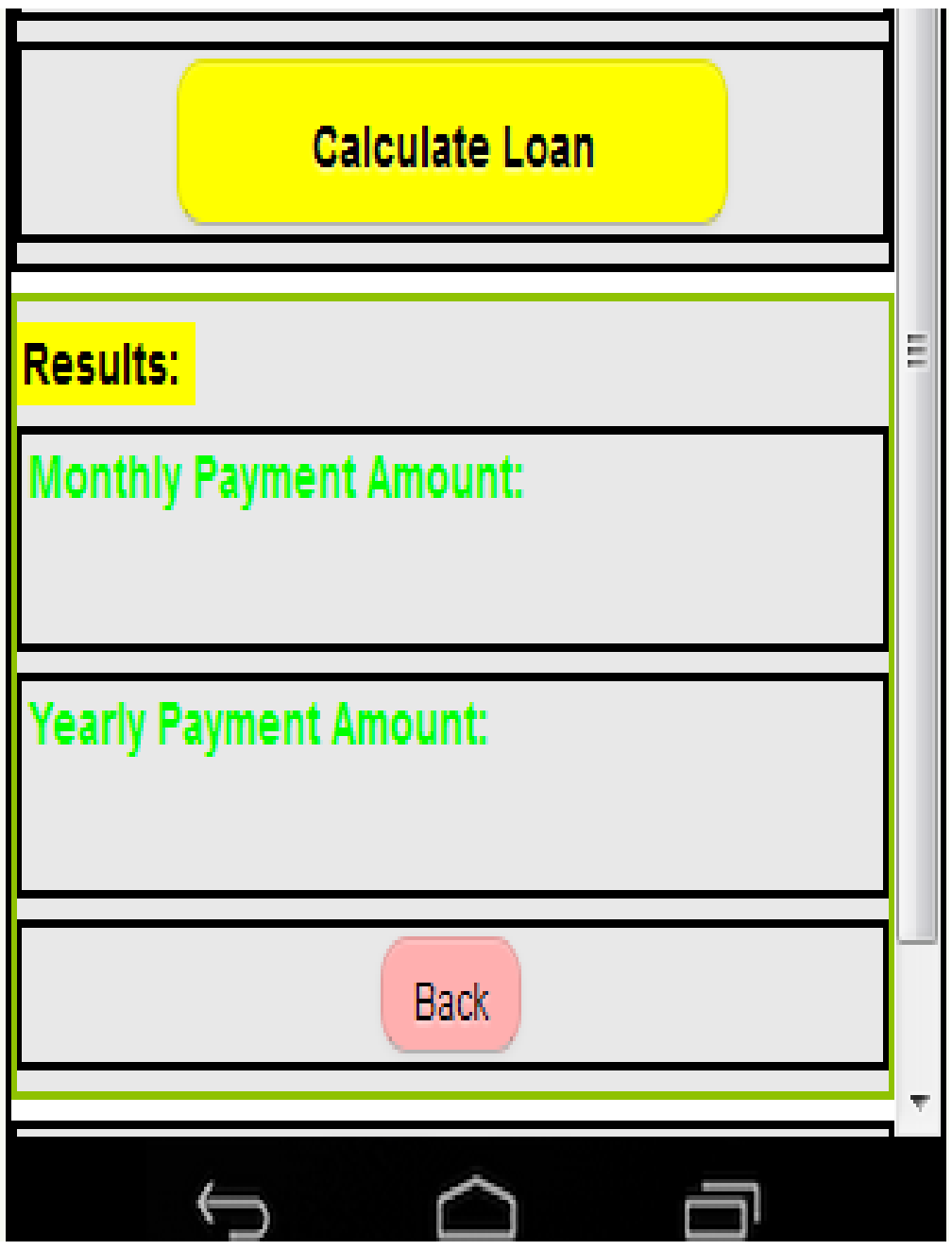

(b)

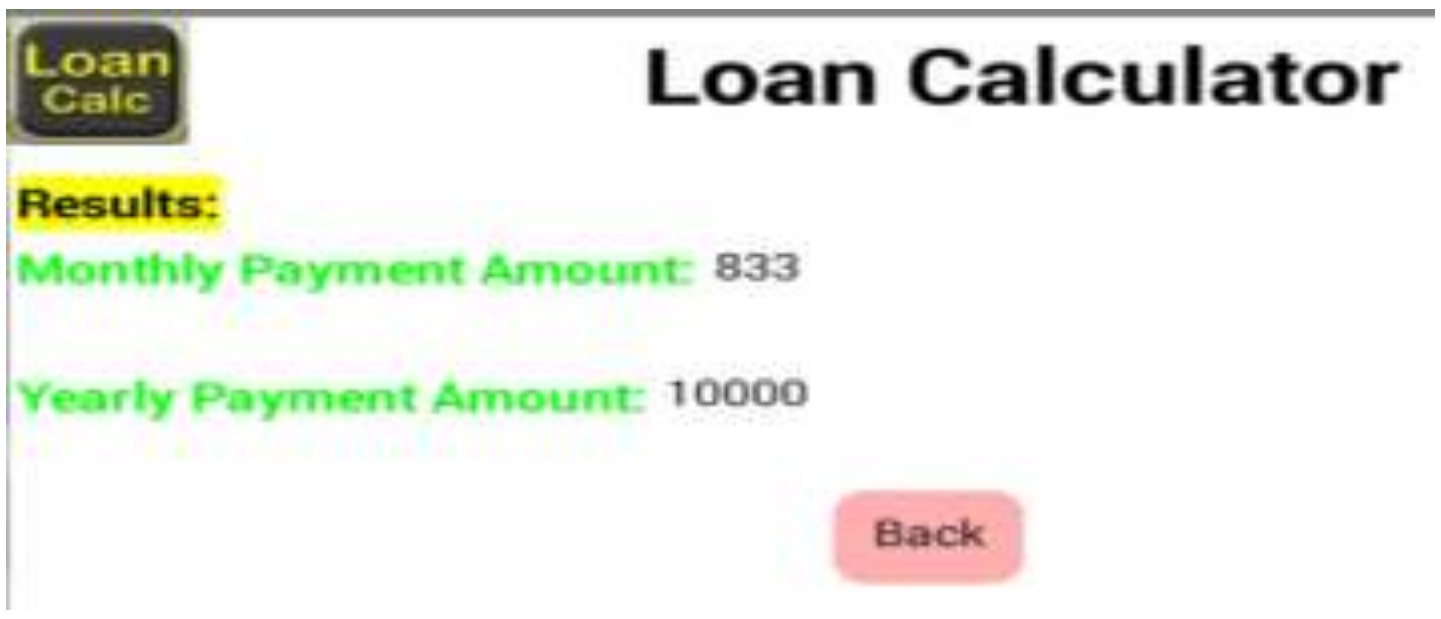

(c)

**Fig. 12.** Loan Calculation module screen (a) input1 (b) input2 (c) output

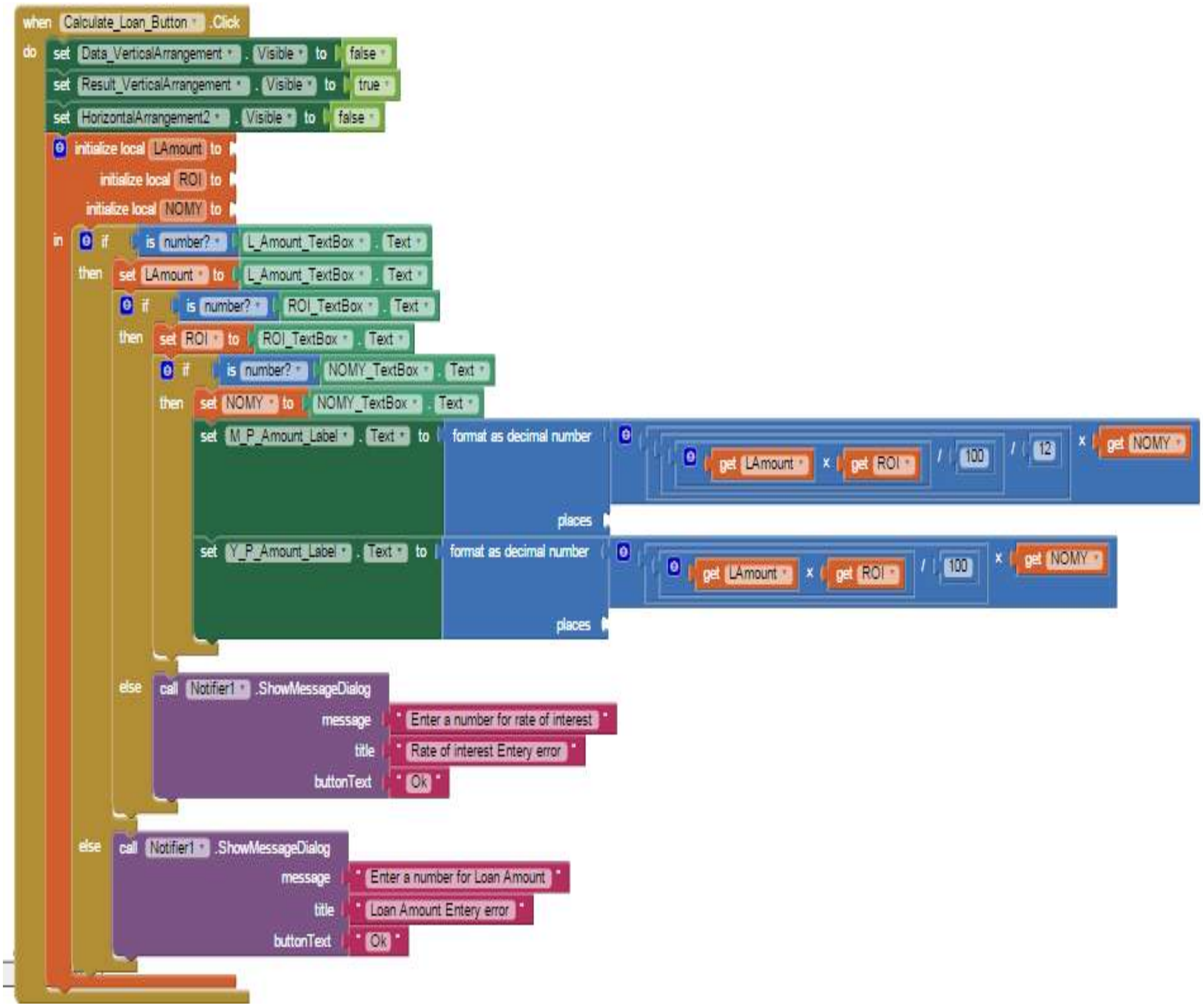

**Fig. 13.** Code block for input and output of loan calculation module

## 4. Evaluation

To evaluate the effectiveness of proposed system, we circulated a web-based questionnaire (Fig. 14) and asked multiple choice questions from end users, and using the user's feedback, we analyzed our system.

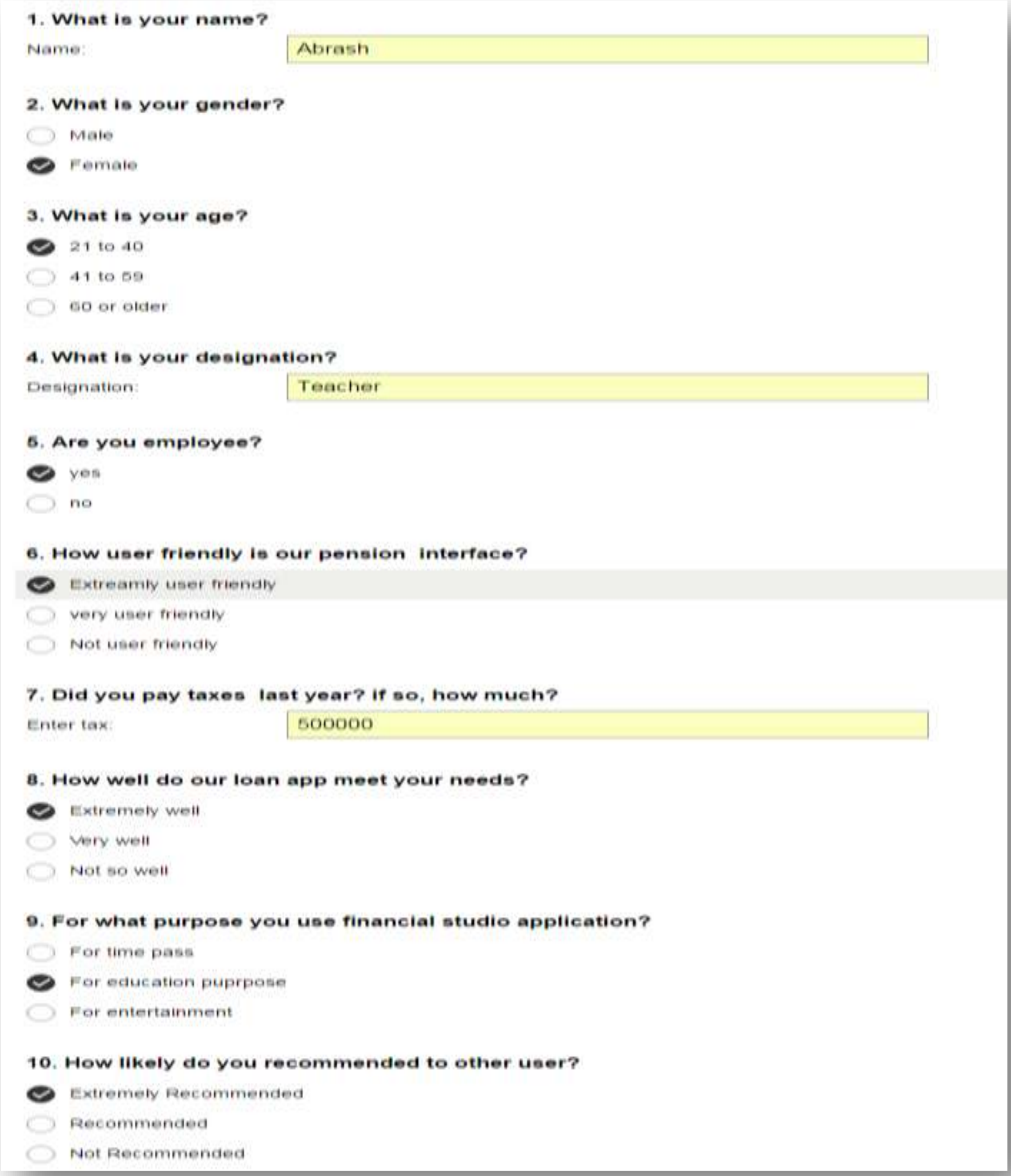

**1. What is your name?**

Name: Abrash

**2. What is your gender?**

Male

[✓ Female]

**3. What is your age?**

[✓ 21 to 40]

41 to 59

60 or older

**4. What is your designation?**

Designation: Teacher

**5. Are you employee?**

[✓ yes]

no

**6. How user friendly is our pension interface?**

[✓ Extreamly user friendly]

very user friendly

Not user friendly

**7. Did you pay taxes last year? if so, how much?**

Enter tax: 500000

**8. How well do our loan app meet your needs?**

[✓ Extremely well]

Very well

Not so well

**9. For what purpose you use financial studio application?**

For time pass

[✓ For education puprpose]

For entertainment

**10. How likely do you recommended to other user?**

[✓ Extremely Recommended]

Recommended

Not Recommended

**Fig. 14.** Web-based Questionnaire

### 4.1 Gender

To enquire about the gender of the respondents, we provide two options: male and female. In response, 32 respondents participated, and out of which, 50% were male and 50% were female (Fig. 15).

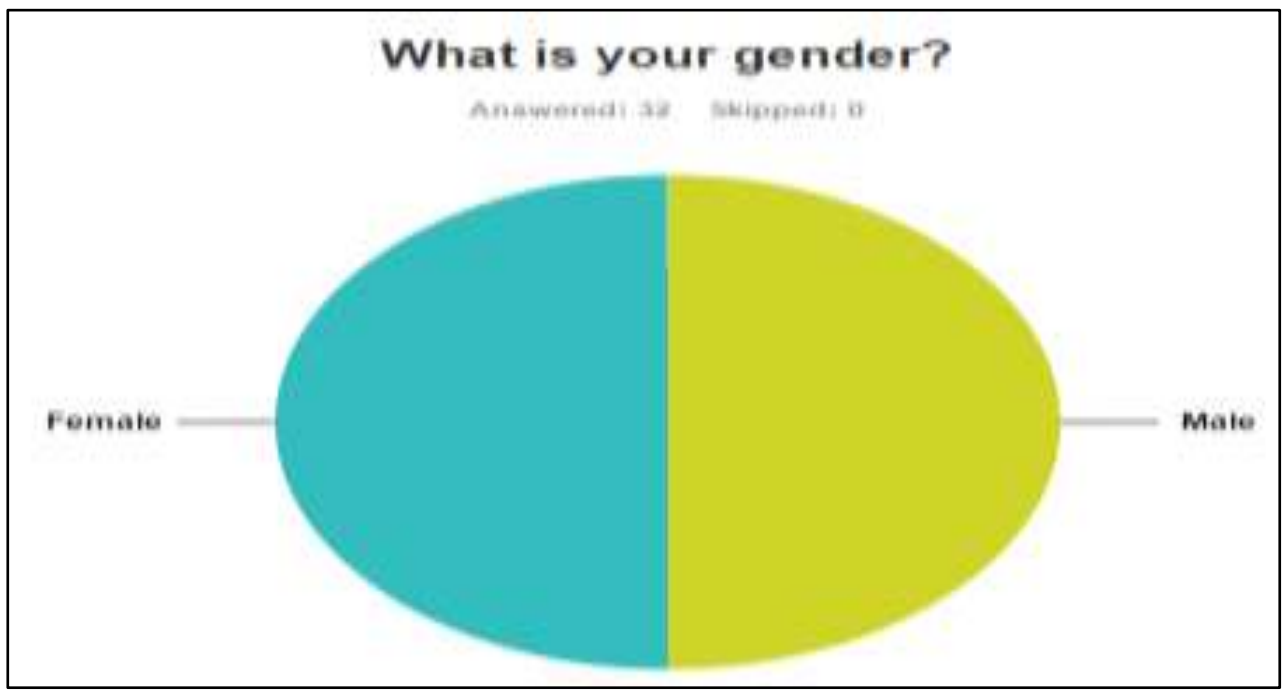

**Fig. 15.** Pie chart of respondents with respect to gender

**Table 1.** Gender-wise descriptive statistics.

| Basic Statistics | | | | |
|---|---|---|---|---|
| **Minimum** 1.00 | **Maximum** 2.00 | **Median** 1.50 | **Mean** 1.50 | **Standard Deviation** 0.50 |

In Table 1, minimum and maximum means the smallest and largest number answer choice that collects not less than one response. It is useful to find the range of answer by subtracting the minimum and maximum. The minimum(1) and maximum(2) presents that there were 16 responses answer(i.e. Male ) and 16 responses answer(i.e. female).The answer choice of all responses shows a median, means and standard deviation. The median of 1.50 (Equal to mean) show that there are equal respondents Female and Male. The mean gives the average of entire responses by adding all number answer choices and then divide them by total amount of number. In this case, a mean of 1.50 represents the overall respondents came in somewhere between Male**,** and the Female. Finally, the standard deviation shows the growth or alteration of your responses, so here the standard deviation is 0.50.

### 4.2 Age

The Fig. 16 shows that there were 40.63% respondents whose age ranges from 21 to 40, 50.00% respondents whose age is between 41 to 59, 9.38% respondents whose age is 60 or older and 2.78% respondents whose age ranges from 55 to 64.

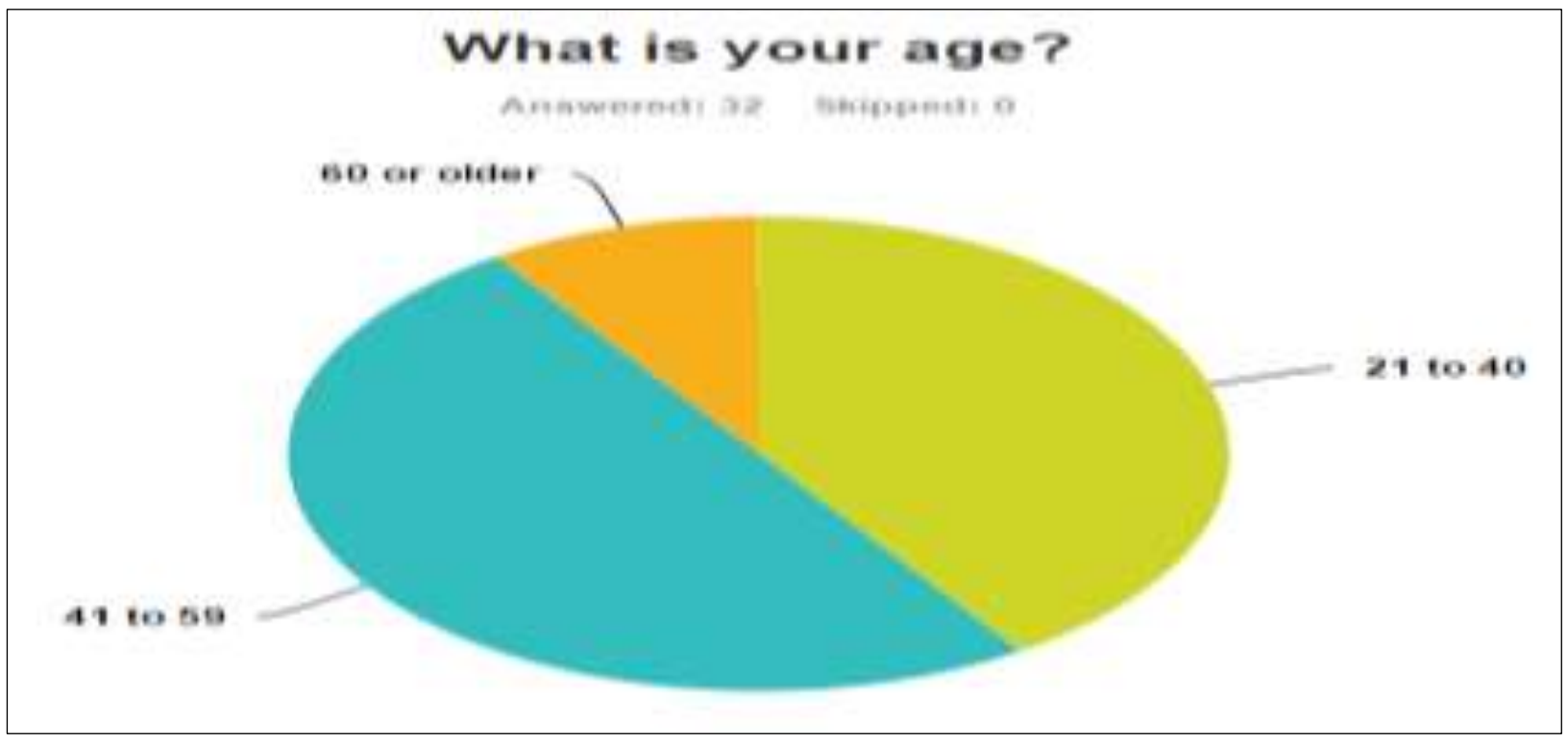

**Fig. 16** Pie chart of respondents with respect to age

**Table 2.** Age-wise Descriptive Statistics

| **Basic Statistics** | | | | |
|---|---|---|---|---|
| **Minimum** 1.00 | **Maximum** 3.00 | **Median** 2.00 | **Mean** 1.69 | **Standard Deviation** 0.63 |

In Table2, minimum (1) and maximum (3) presents that there were 13 responses in the uppermost answer (i.e. age 21 to 40) and 3 responses in the lowermost answer (i.e. age 60 or older). The median of 2.00 (higher than the 1.69 mean) show that there were more respondents who were in age (41 to 59) than respondents who were in age (21 to 40). A mean of 1.69 shows that overall respondents came in somewhere between age (21 to 40), and the age (41 to 59). Finally, the standard deviation shows the growth or alteration of your responses**,** so here the standard deviation is 0.63.

**4.3 Employee**

The aim of this question is to know that whether the person using our app is employee or not? Fig.17 shows that out of 32 respondents, 93.75% were employees 6.25% were non-employees using our app.

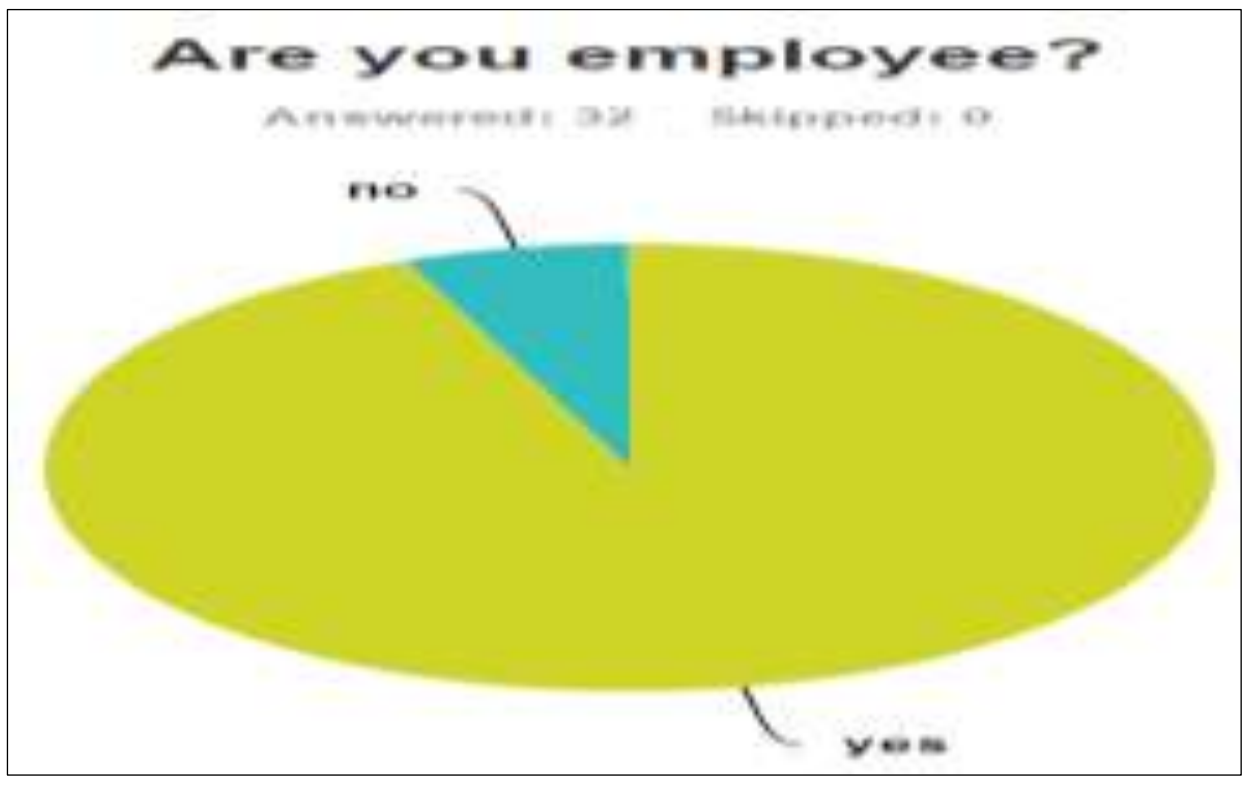

**Fig. 17** Pie chart of respondents with respect to employee

**Table 3.** Descriptive statistics about employee or not?

| **Basic Statistics** | | | | |
|---|---|---|---|---|
| **Minimum** 1.00 | **Maximum** 2.00 | **Median** 1.00 | **Mean** 1.06 | **Standard Deviation** 0.24 |

In Table 3, minimum (1) and maximum (2) presents that there were 30 responses in the uppermost answer (i.e. Yes) and 2 responses in the lowermost answer (i-e No). The median of 1.00 (less than the 1.06 mean) show that there were more respondents who are employees. In this case, a mean of 1.06 shows that overall respondents came in somewhere between Yes and No. The mean gives the average of entire responses. Finally, the standard deviation shows the growth or alteration of your responses**,** so here the standard deviation is 0.24.

**4.4 User Friendly**

In this question we want to know the opinion of user about the user friendliness of our pension apps. In Fig. 18, we observe that 54.84% of the respondents found the application as "*Extremely user friendly*", 25.81% of the respondents were of the view that the application is "*very user friendly*", and 19.35% of respondents declared it as "*not user friendly*".

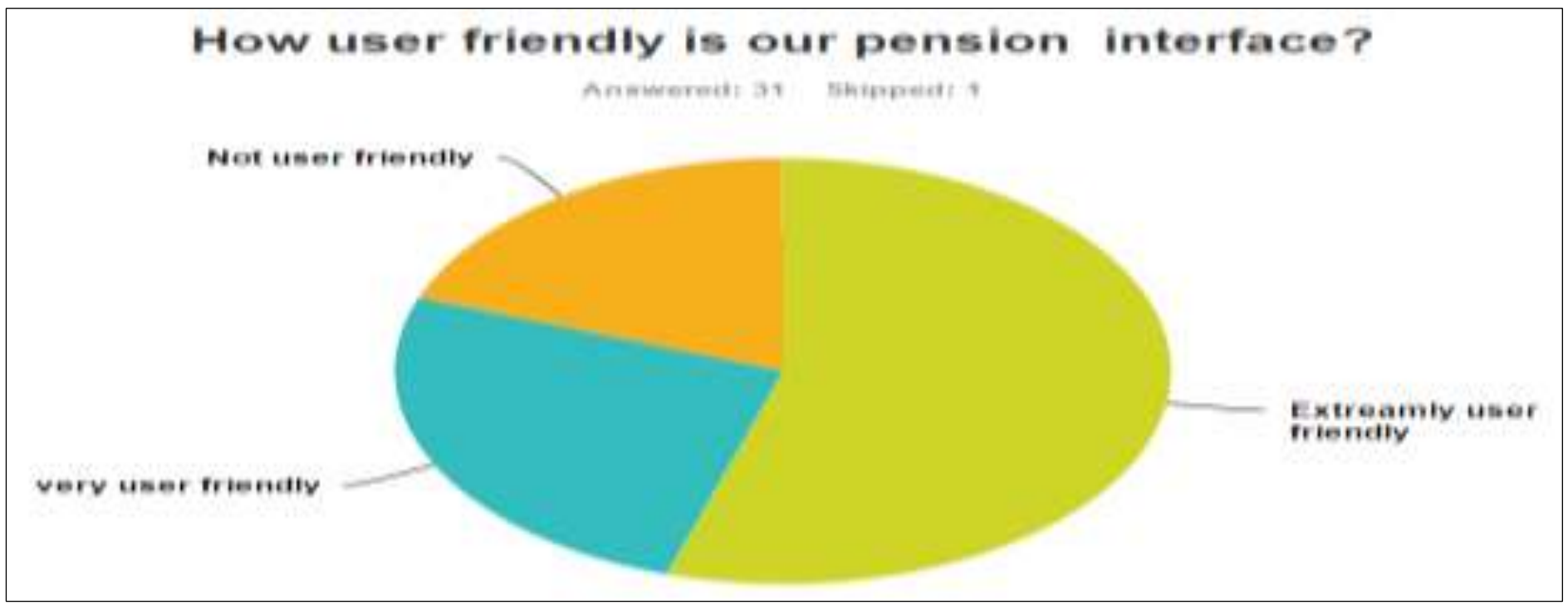

Figure 18. Pie Chart about user friendly pension interface

**Table 4.** Descriptive statistics about user friendliness of the application

| **Basic Statistics** | | | | |
|---|---|---|---|---|
| **Minimum** 1.00 | **Maximum** 3.00 | **Median** 1.00 | **Mean** 1.65 | **Standard Deviation** 0.78 |

In Table 4, minimum (1) and maximum (3) presents that there were 17 responses in the uppermost answer (i.e. Extremely user friendly ) and 6 responses in the lowermost answer(i.e. not user friendly). The median of 1.00 (lower than the 1.65 mean) show that there were some respondents who said pension interface is very user friendly for them. The mean gives the average of entire responses. Finally, the standard deviation shows the growth or alteration of your responses**,** so here the standard deviation is 0.78.

**4.5 Loan app meet user needs?**

In this question we want to investigate about the usefulness of our loan app. Fig. 19 shows that 56.25% of the respondents found it as "*Extremely well*", 25.00% said "*very well*", and 18.75% declared it as "*Not so well*".

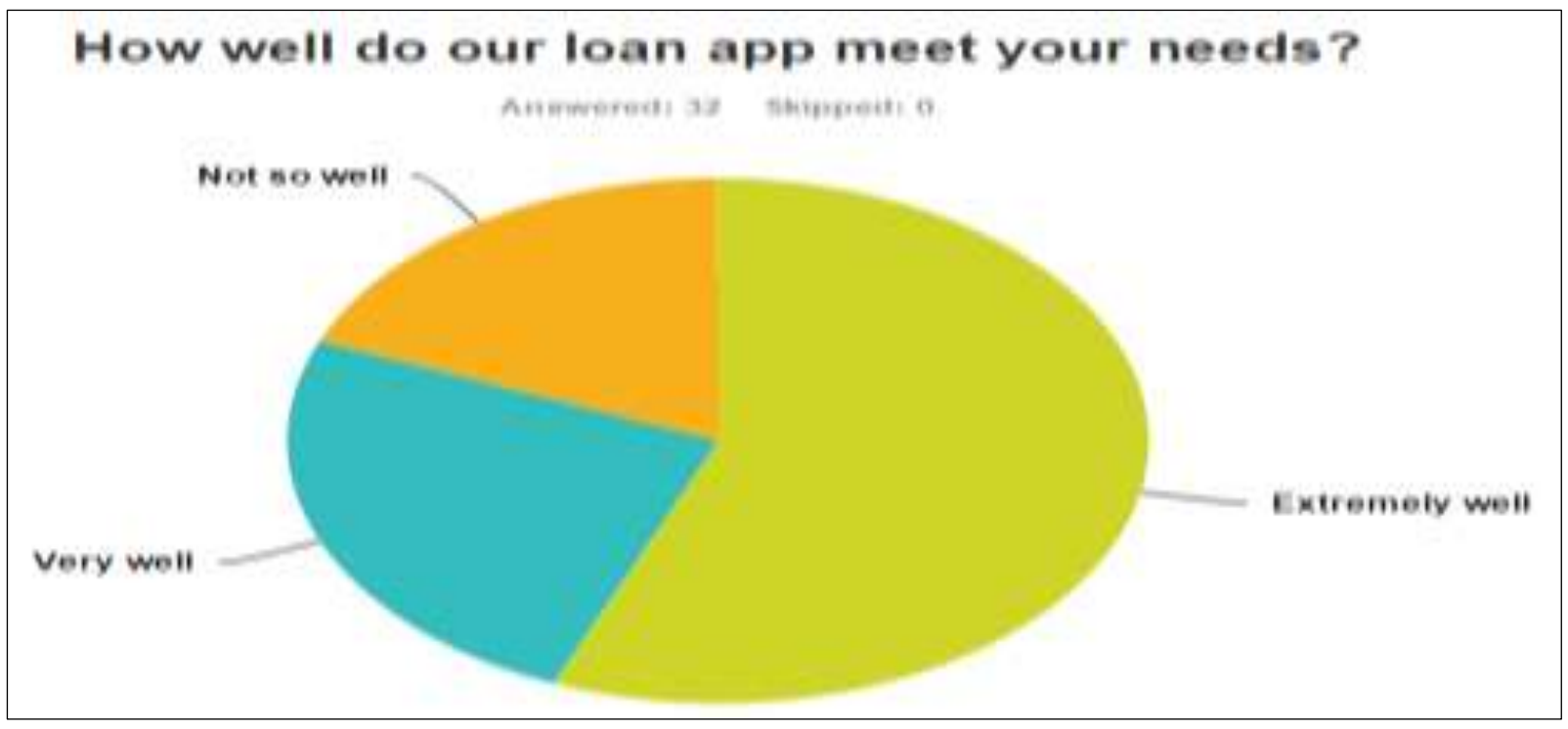

**Fig. 19** Pie Chart for loan app usefulness

**Table 5.** Descriptive statistics about the usefulness of the loan app

| **Basic Statistics** | | | | |
|---|---|---|---|---|
| **Minimum** 1.00 | **Maximum** 3.00 | **Median** 1.00 | **Mean** 1.63 | **Standard Deviation** 0.78 |

In Table 5 minimum (1) and maximum (3) presents that there were 18 responses in the uppermost answer (i.e. extremely well) and 6 responses in the lowermost answer (i.e. Not so well). The median of 1.00 (lower than the 1.63 mean) show that there were some (i.e. 8) respondents who said that loan app is very well. The mean gives the average of entire responses. Finally, the standard deviation shows the growth or alteration of your responses**,** so here the standard deviation is 0.78.

**4.6 Objective of financial studio**

This question intends to ask from user about why he/she uses this application. Fig. 20 shows that 9.68% of the respondents use it for "*entertainment*" response, 80.65% used for educational purpose and 9.68% used it for "*time passing*".

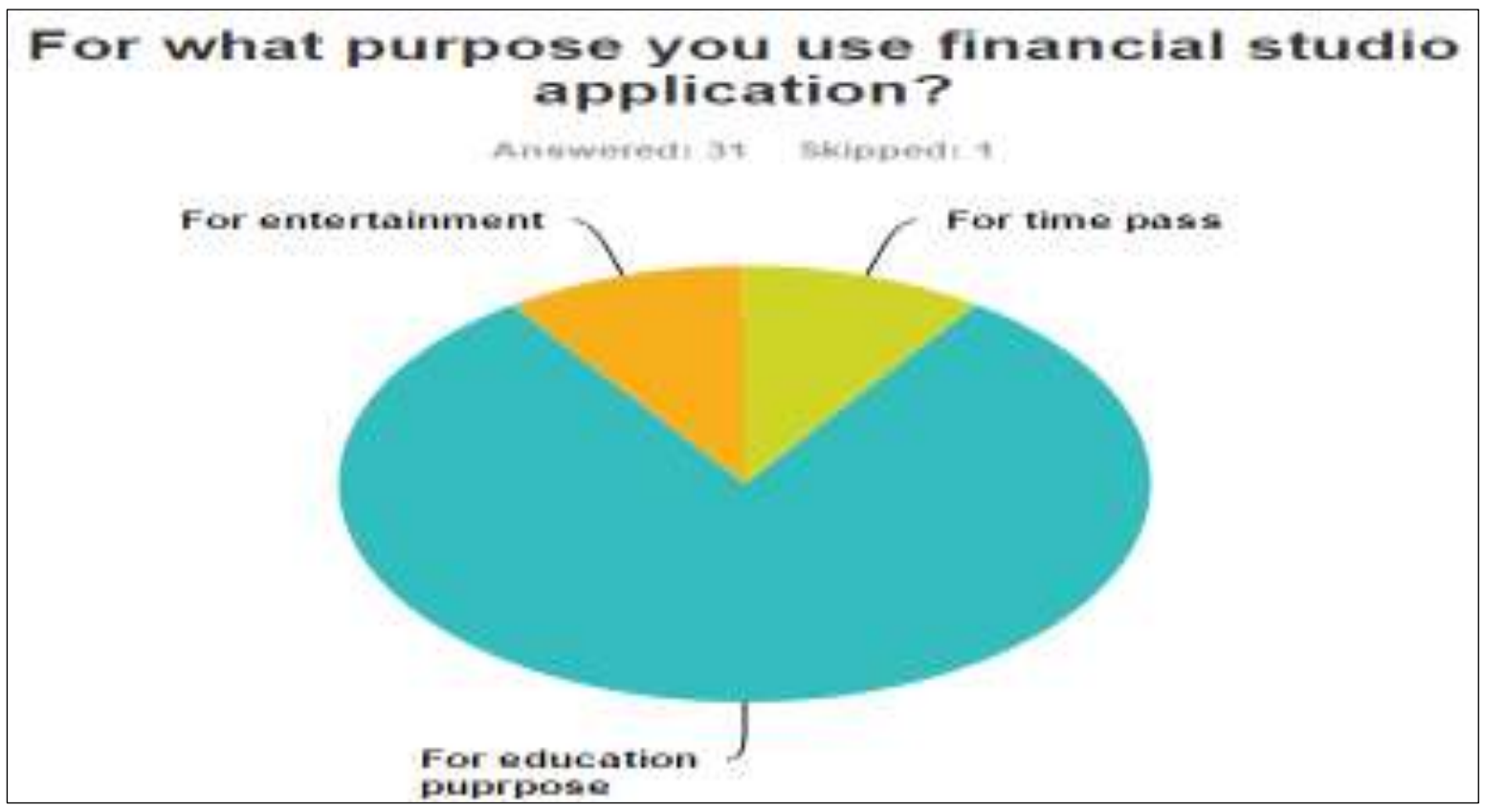

**Fig. 20** Pie Chart about objective of financial studio

**Table 6.** Showing what purpose financial application use Basic Statistics.

| **Basic Statistics** | | | | |
|---|---|---|---|---|
| **Minimum** 1.00 | **Maximum** 3.00 | **Median** 2.00 | **Mean** 2.00 | **Standard Deviation** 0.44 |

In the Table 6, minimum (1) and maximum (3) presents that there were 3 responses in the uppermost answer (i.e. For time pass) and 3 responses in the lowermost answer (i.e. For entertainment). The median of 2.00 (equal to mean) show that there were more respondents who use this app for education purpose Finally, the standard deviation shows the growth or alteration of your responses, so here the standard deviation is 0.44.

**4.7 Do you recommend this app?**

The Fig. 21 shows that 53.13% of the respondents extremely recommended, 28.13% of the respondents agree to recommend our app, and 18.75% of the respondents have not recommended the interface of application to others.

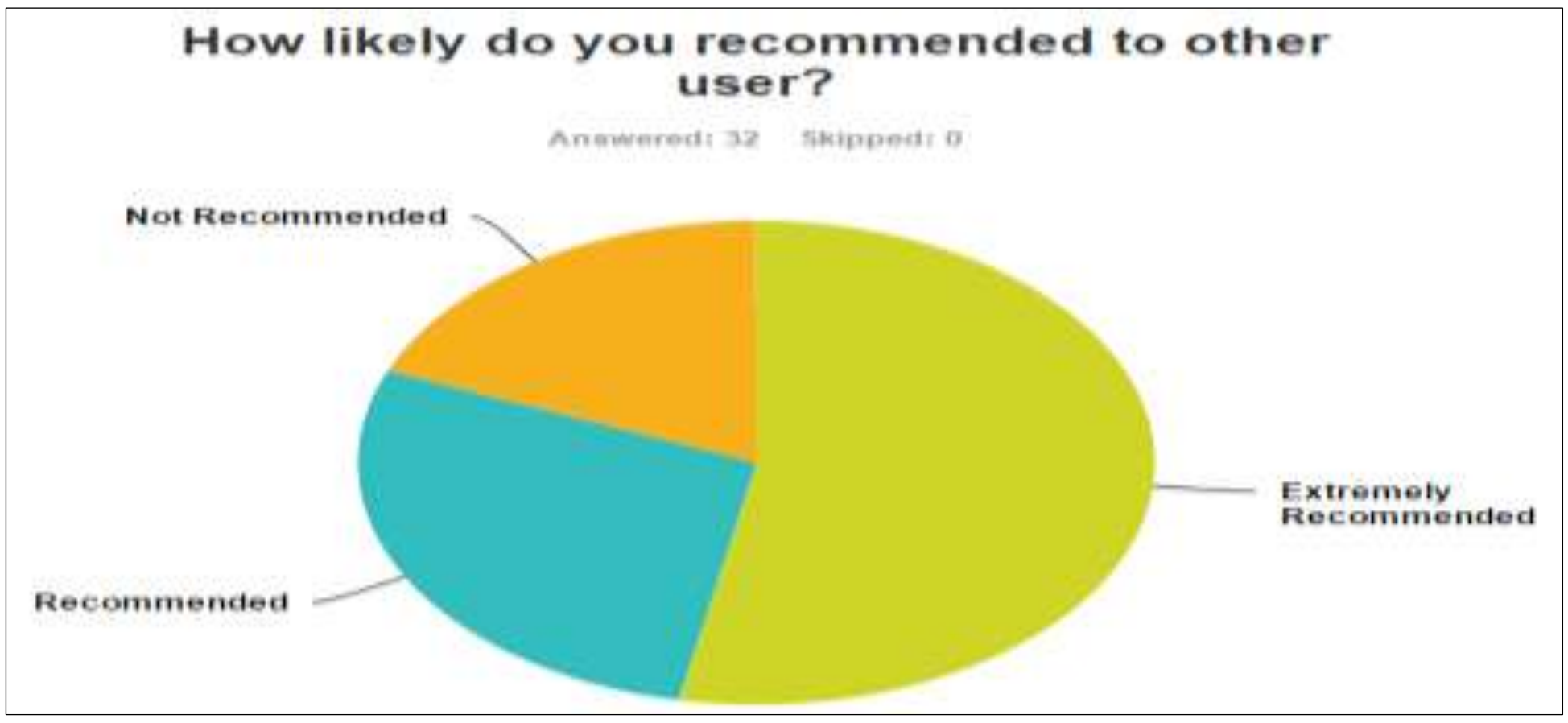

**Fig. 21** Pie chart showing the recommendation of the application

**Table 7.** Descriptive statistics about how likely do you recommended to the others

| **Basic Statistics** | | | | |
|---|---|---|---|---|
| **Minimum** 1.00 | **Maximum** 3.00 | **Median** 1.00 | **Mean** 1.66 | **Standard Deviation** 0.77 |

In Table7, minimum (1) and maximum (3) presents that there were 17 responses in the uppermost answer (i.e. extremely recommended) and 6 responses in the lowermost answer (i.e. Not recommended). The median of 1.00 (lower than the 1.66 mean) some respondents who agreed to recommend our app than respondents who were not recommended. In this case, a mean of 1.66 shows that overall respondents came in somewhere between extremely recommended, and the recommended. Finally, the standard deviation shows the growth or alteration of your responses, so here the standard deviation is 0.77.

## 4. Conclusion and Future work

This work deals with the development of android based financial studio, an integrated application for calculating tax, pension, zakat, and loan. It has four modules namely, (i) tax calculation module, (ii) pension calculation module, (iii) Zakat calculation module, and (iv) loan calculation module.

The tax calculation module is used to perform tax related calculations required by a salaried person in government and non-govt. organization. The tax rules and rates are based on government of Pakistan (2014-15) regulations. The pension calculation module is used to perform calculations pertaining to pension related issues of a retired person of some organization. The application assists in computing total gratuity, net pension, increases on pension, medical allowances, and total monthly pension. The third module is related to zakat calculations payable on gold, silver, cash, business, and property. Finally the loan calculation module is used to perform loan related calculations on the basis of input provided by the user. The proposed system has shown an efficient performance with respect to user friendly and interactive design by using app inventor technology. The statistical analysis shows that users have found the application as beneficial for Financial calculations.

***Future directions:*** Following are the possible future directions: (i) To incorporate currency convertor module, (ii) to extend the current pension calculation module by incorporating pension calculation related to death cases, (iii) to enhance the existing tax calculation module to incorporate tax calculations for business and property.

## 5. Acknowledgment

In the name of Allah, the Most Gracious and the Most Merciful, Alhamdulillah, all praises t Allah for the strengths and His blessing in completing this research paper. We would like to express our deepest gratitude to our Supervisor, Dr. Muhammad Zubair Asghar, for his excellent guidance, caring, patience, and providing us with an excellent atmosphere for doing our research work.